\documentclass[twocolumn,preprintnumbers,amsmath,amssymb,floatfix]{revtex4}

\usepackage{graphicx}
\usepackage{indentfirst}
\usepackage{physics}
\usepackage{caption}
\usepackage{wrapfig}
\usepackage{subfig}
\usepackage{listings}
\usepackage{float}
\usepackage{amsmath}
\begin{document}
\author{Ashod Khederlarian$^{1}\footnotemark[2]$, Carmen Almasri$^{2}\footnote[2]{These authors are of equal contributions}$, and Leonid Klushin$^{1}$ }
\affiliation{$^1$Physics Department,  American University of Beirut, Beirut 1107 2020, Lebanon}
\affiliation{$^2$Department of Physics and Astronomy, University of California, Irvine, CA 92697, USA}

\date{\today}
\begin{abstract}
We explore semi-dilute and concentrated oligomers and polymers in a broad range of polymerization indices N ranging from 1 to a 100 and in a range of monomer number densities $\phi$ from 0.1 to 0.8 via molecular dynamics simulations and under good solvent conditions. This parameter range covers both no-overlap and strong chain overlap regimes, as quantified by the polymer packing fraction $0.1\le\Phi\le{14}$. Contrary to some common beliefs, the non-ideal part of the osmotic pressure demonstrates strong finite size effects. In the overlap regime, it deviates substantially from the scaling form of de Cloizeaux. The finite size correction term is proportional to 1/N, irrespective of $\Phi$. We propose a simple phenomenological description of the osmotic pressure in the infinite chain limit and of the monomer density dependence of the 1/N correction term. We extend the treatment of finite size effects to cover binary mixtures with 2 different chain lengths, and demonstrate that the proposed equation of state is applicable with an effective mass-averaged inverse chain length $1/N_{eff}$. We also discuss finite size effects in the density dependence of the gyration radius. 
\end{abstract}
\title{Molecular Dynamics Simulations of Semi-Dilute and Concentrated Solutions: Unexpected Finite Size Effects in Osmotic Pressure}
\maketitle
\section{Introduction}

\section{Model}
First, we consider a polymer chain solution confined in a box $L_x=L_y=L_z=L$ with periodic boundary conditions in all directions. In all of our studies, $L$ is of the order 10 units; large enough to avoid self interaction through the periodic boundaries for the chain lengths considered.

For interactions, we used the standard bead-spring model of Kremer and Grest\cite{kremer1990dynamics}{,} where bonded beads interact with the spring potential $U_{FENE}(r)=-0.5kR^2\ln[1-({r}/{R})^2]$, and pairs of beads interact with the repulsive part of the Lennard Jones potential, $U_{LJ}(r)=4{\epsilon}[(\sigma/r)^{12}-(\sigma/r)^{6}]+\epsilon$ for $r<2^{1/6}\sigma$ and 0 otherwise. Here r is the radial distance between beads, $\sigma=1$ serves as the unit of length, $\epsilon=1$ in terms of the thermal energy $K_bT=1$, $R=1.5\sigma$, $k=30\epsilon{\sigma^{-2}}$.   

\section{MD simulation}
The forward integration in time was driven by the usual Langevin equation including a thermostat \cite{allen2017computer}
\begin{equation}
m\frac{d\boldsymbol{v}}{dt}=\boldsymbol{f}_c-\gamma{\boldsymbol{v}}+\eta(t)
\end{equation}
where $\boldsymbol{f}$ is the net force on a bead from the potentials, $\boldsymbol{v}$ is its velocity, $\gamma=0.25$ is the drag coefficient, and $\eta$ the random force due to density fluctuations in the solvent. The equation was solved using the velocity-verlet algorithm with a time step $dt=5\times10^{-4}\tau$ $(\tau=\sqrt{m\sigma^2/\epsilon}=1)$.
Initially, the system is setup and equilibrated (for $2\times10^6dt$) using HOOMD-blue \cite{anderson2020hoomd}{.} Then, the equilibrated configuration was used to start another simulation, where every $100dt$ (to make sure samples are uncorrelated), the local pressure tensor was calculated according to the Irving-Kirkwood expression\cite{irving1950statistical}
\begin{equation}
P_{\alpha\beta}=\frac{1}{V}\sum_{i}\frac{p_i^{\alpha}p_i^{\beta}}{m}-\frac{1}{V}\sum_{ij}f_{ij}\frac{r_{ij}^\alpha}{r_{ij}}r_{ij}^\beta\int_{C_{ij}}d\lambda
\end{equation}  
After the simulation, this tensor was averaged over all samples, and the ensemble averaged osmotic pressure of the system was obtained from the diagonal component of the tensor that corresponds to the axis along which the slicing for $\lambda$ was done.

\section{Results}
\subsection{WCA fluid}
As a check for the pressure tensor algorithm, the well established results for a gas of purely repulsive monomers was reproduced. An equation of state is given in \cite{hess1998thermomechanical} which is of the form
\begin{equation}
P=\phi{k_B}T+P_{WCA}
\end{equation}
Where P is the total pressure of the system, and $\phi$ is the monomer number density, with the second term given by
\begin{equation}
P_{WCA}=\phi{k_B}{T}\bigg(\frac{\phi{B^{WCA}}}{(1-\phi{v_{eff}})^2}+2\frac{(\phi{v_{eff}})^2}{(1-\phi{v_{eff}})^3}\bigg)
\end{equation}
where the virial coefficient $B^{WCA}\approx2.2$ and the temperature dependent effective volume is evaluated from $v_{eff}(T)=(\pi/6)d_{eff}^3$, where $d_{eff}$ is defined as the separation between monomers over which their interaction potential is equal to $k_BT$. For $k_BT=1$, $v_{eff}\approx0.525$.

\begin{figure}[H]
	\centering
	\includegraphics[width=1\linewidth]{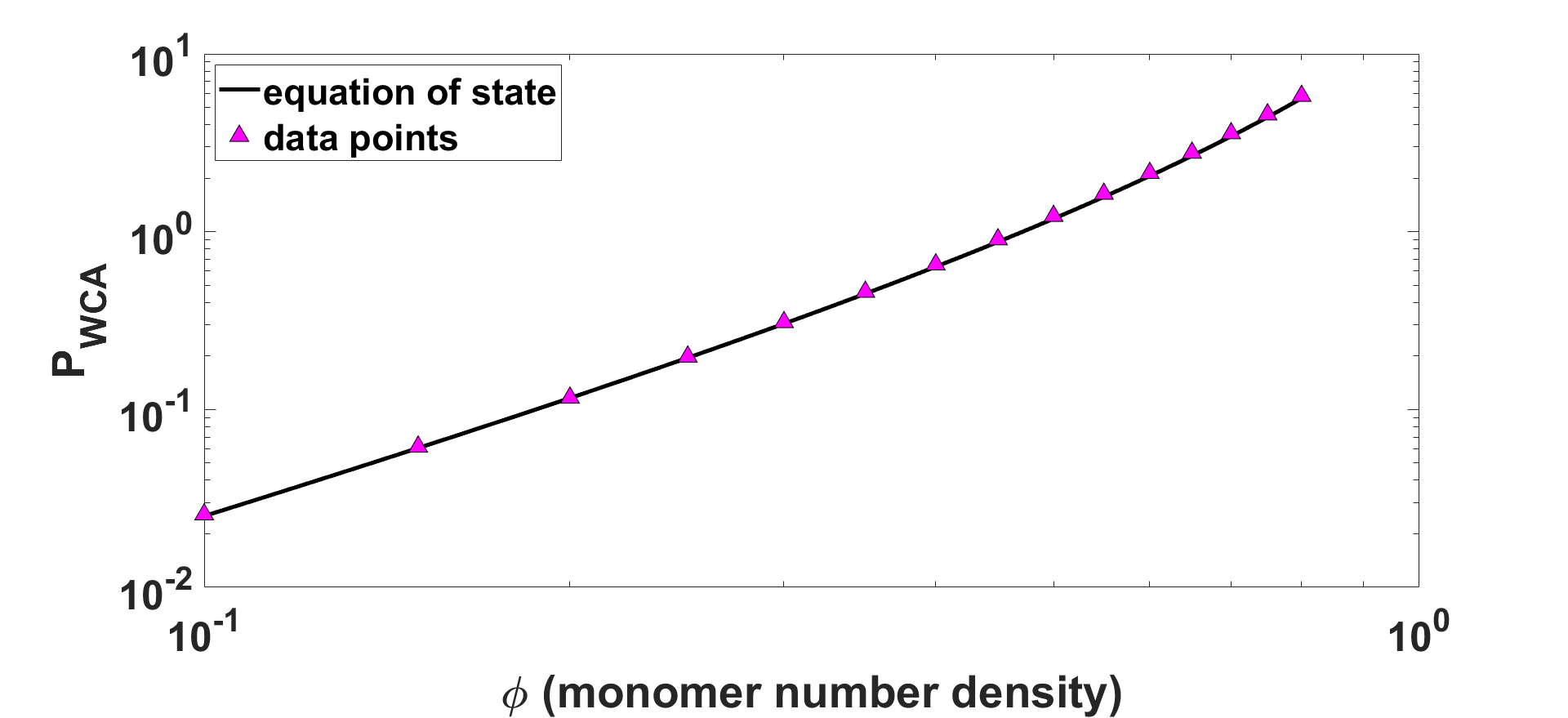}
	\caption{$P_{WCA}$ vs. monomer number density. Pink triangles are data points obtained from our simulations, while the line is the well established equation of state.}
	\label{fig:Fig11}
\end{figure}

To evaluate the pressure numerically, the system was initially equilibrated for $10^6$ MD steps, and afterwards a sample was obtained for the pressure tensor every 100 steps, up to $100,000$ samples. The results are compared to the above equation of state in figure \ref{fig:Fig11}.
\subsection{Monodisperse Solution}
\begin{figure}[H]
	\centering
	\includegraphics[width=1\linewidth]{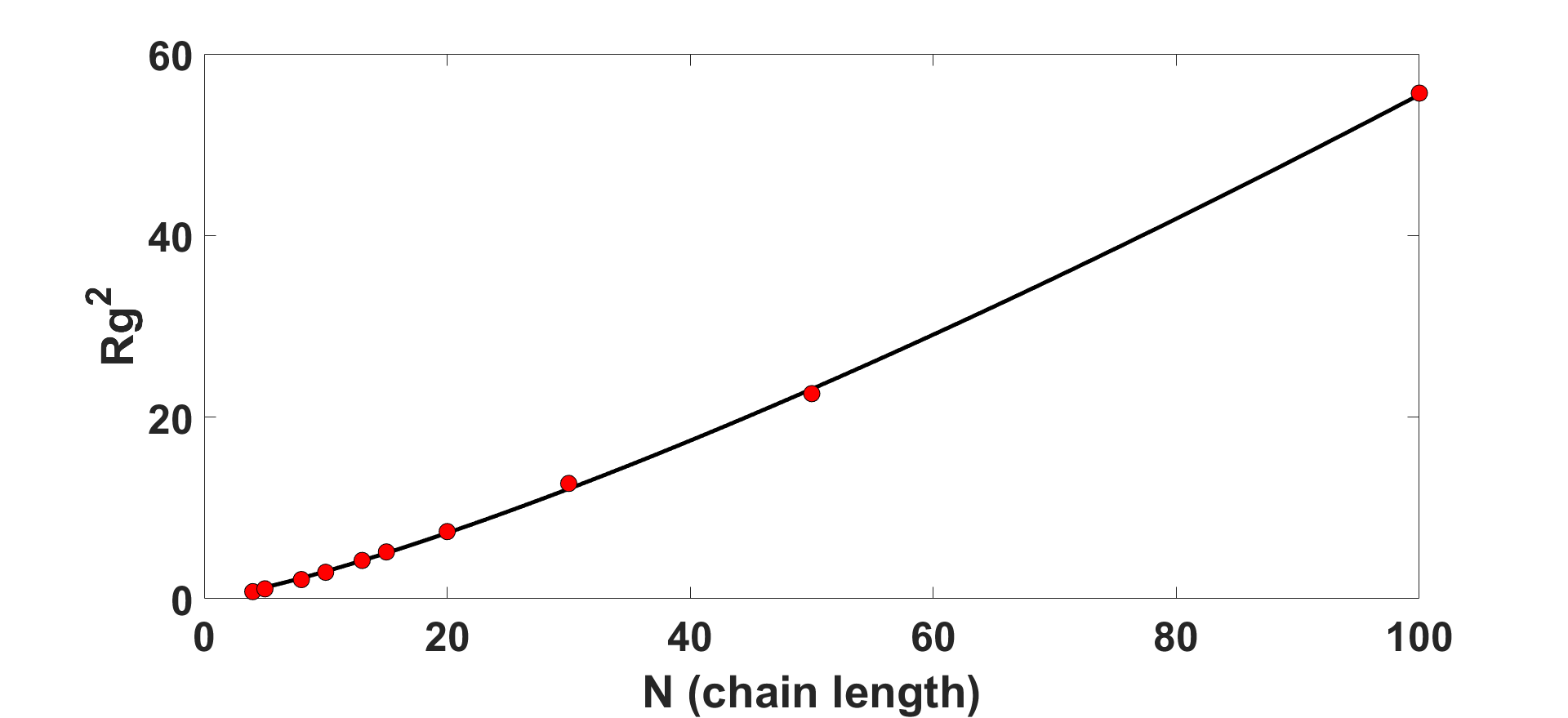}
	\caption{The square of the gyration radius as a function of chain length in the limit of 0 density. The fit goes like $N^{1.265}$.}
	\label{fig:Fig8}
\end{figure}

For Monodisperse solutions, gyration radii were studied as a function of monomer density and chain length. It is defined as
\begin{equation}
\vec{R_g}^2=\frac{1}{N}\sum_{i}(\vec{r_i}-\vec{R_{cm}})^2	
\end{equation}
N being the length of the chain, $\vec{R_{cm}}$ the center of mass position vector of the chain, and the sum is over the monomers within a single chain. The 0 density limit was obtained by placing a single chain of size $N$ in a box of size $L=10N$, allowing it to equilibrate for 
$2.5\times10^6$ MD steps, and then sampling every 100 for a total of $25,000$ samples. The data is shown in figure \ref{fig:Fig8}, along with a power fit That is fairly close to the exponent of 1.176 suggested in \cite{caracciolo2006polymer}.
\begin{equation}
R_g^2(0,N)=0.1638N^{1.265}
\end{equation}

\begin{figure}[H]
	\centering
	\includegraphics[width=1\linewidth]{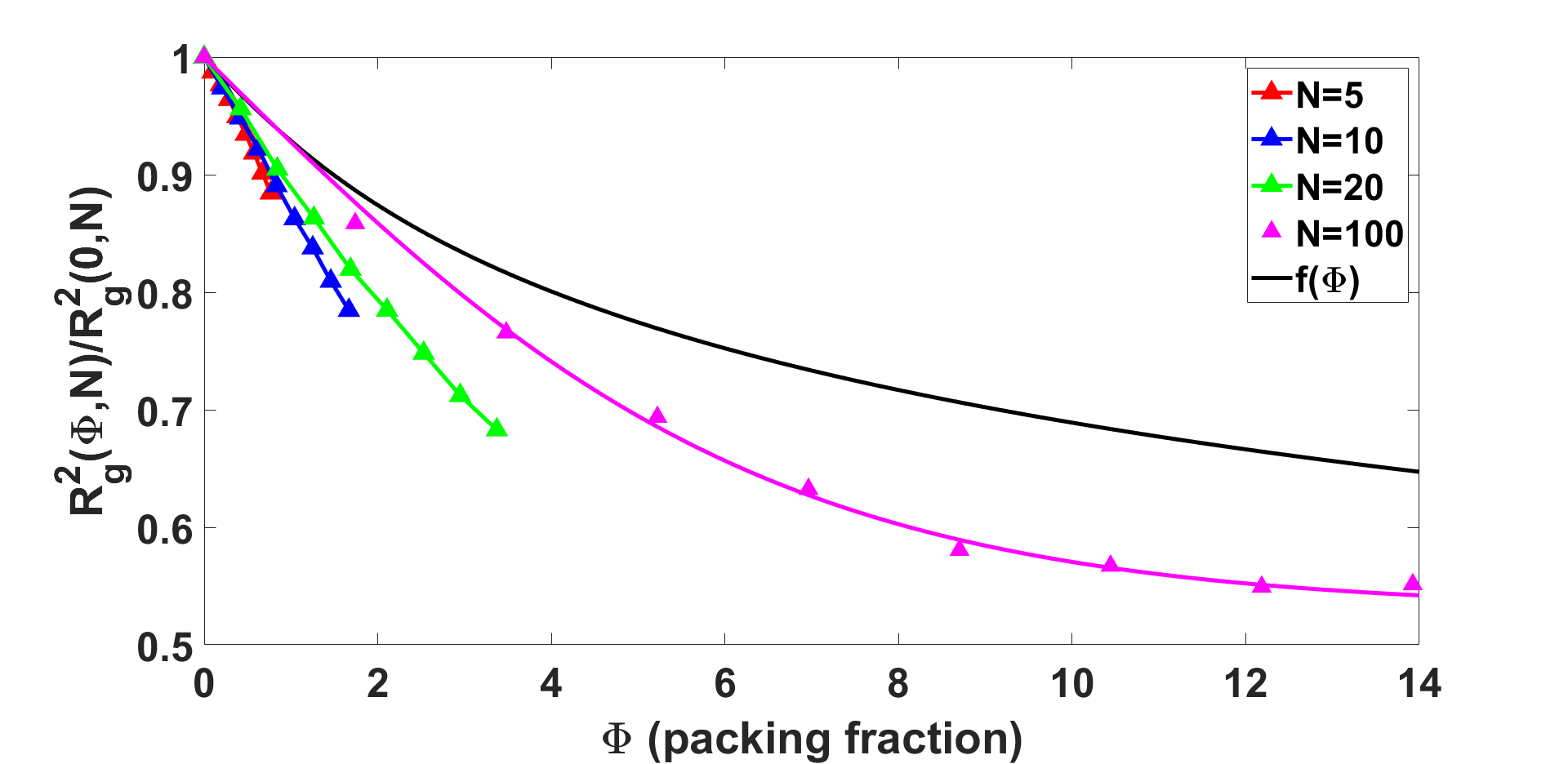}
	\caption{Dependence of normalized gyration radii on packing fraction $\Phi$ for chains of different lengths $N=[5,10,20,100]$. $f(\Phi)$ (black line) is the expected dependence in the infinite chain length limit, according to equation 8.}
	\label{fig:Fig9}
\end{figure}

\begin{figure}[H]
	\centering
	\includegraphics[width=1\linewidth]{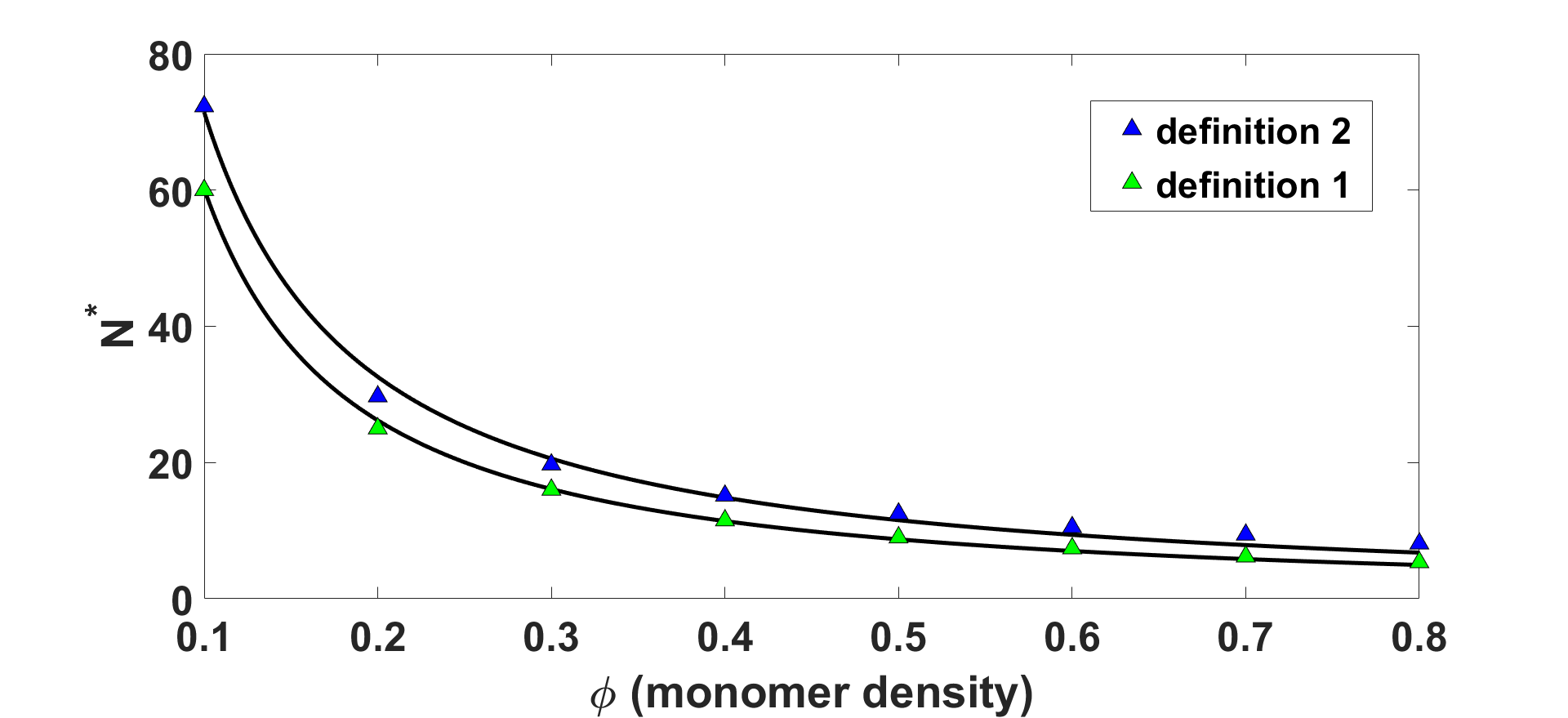}
	\caption{$N^*$ as a function of density with 2 possible definitions that give a range of values below which is the non-overlap regime and above which is the overlap regime.}
	\label{fig:Fig10}
\end{figure}

Increasing the density would naturally reduce the gyration radius due to the repulsive interaction between chains. This effect is typically analyzed by studying how $R_g^2(\Phi,N)/R_g^2(0,N)$ depends on a normalized density called the polymer packing fraction
\begin{equation}
\Phi=\frac{4}{3}\pi{R_g^3(0,N)}\frac{\phi}{N}
\end{equation}
with $\phi$ being the monomer density, not to be confused with capital $\Phi$. In \cite{pelissetto2008osmotic}, a functional form for this dependence is suggested in the infinite chain length limit $f(\Phi)=R_g^2(\Phi,N\rightarrow\infty)/R_g^2(0,N\rightarrow\infty)$, that interpolates between the known behaviors for small and large $\Phi$
\begin{equation}
f(\Phi)=\frac{(1+0.33272\Phi)^{0.115}}{(1+0.986633\Phi+0.499436\Phi^2+0.049597\Phi^3)^{0.115}}
\end{equation}
We compare this with our data for chains of lengths $[5,10,20,100]$ in figure \ref{fig:Fig9}. As expected, finite size corrections are needed in the form of $1/N$, and the points tend to the suggested dependence with increasing chain length.
Lastly, for future use, we define $N^*$, the chain length that separates overlap and non-overlap regimes for a given density, as the value of N for which the following equality holds
\begin{equation}
\frac{N}{\frac{4}{3}\pi{R_g^3(0,N)}}=\phi
\end{equation}
However, it should be noted that this definition is rather arbitrary, and one can equally well use 
\begin{equation}
\frac{N}{\frac{4}{3}\pi{R_g^3(\phi,N)}}=\phi
\end{equation}
We use both to define a 'range' of values at a given density that separate overlap with non-overlap. This is shown in figure \ref{fig:Fig10} along with power-law fits
\begin{equation}
N^*_1=5.261\phi^{-1.133}
\end{equation}
\begin{equation}
N^*_2=3.797\phi^{-1.2}
\end{equation}

\begin{figure}[H]
	\centering
	\includegraphics[width=1\linewidth]{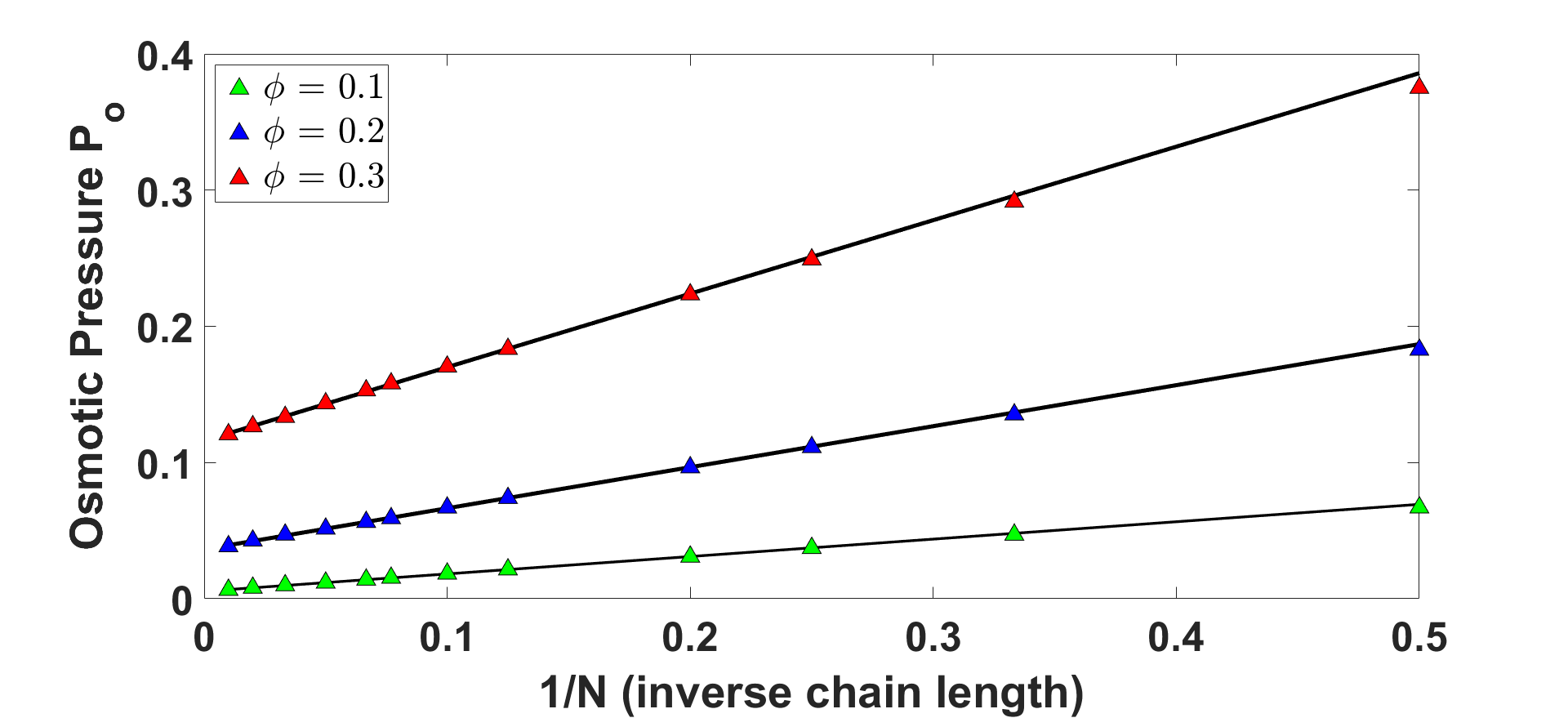}
	\caption{Osmotic pressure vs inverse chain length for densities 0.1 (green), 0.2 (blue), and 0.3 (red).}
	\label{fig:Fig1}
\end{figure}

\begin{figure}[H]
	\centering
	\includegraphics[width=1\linewidth]{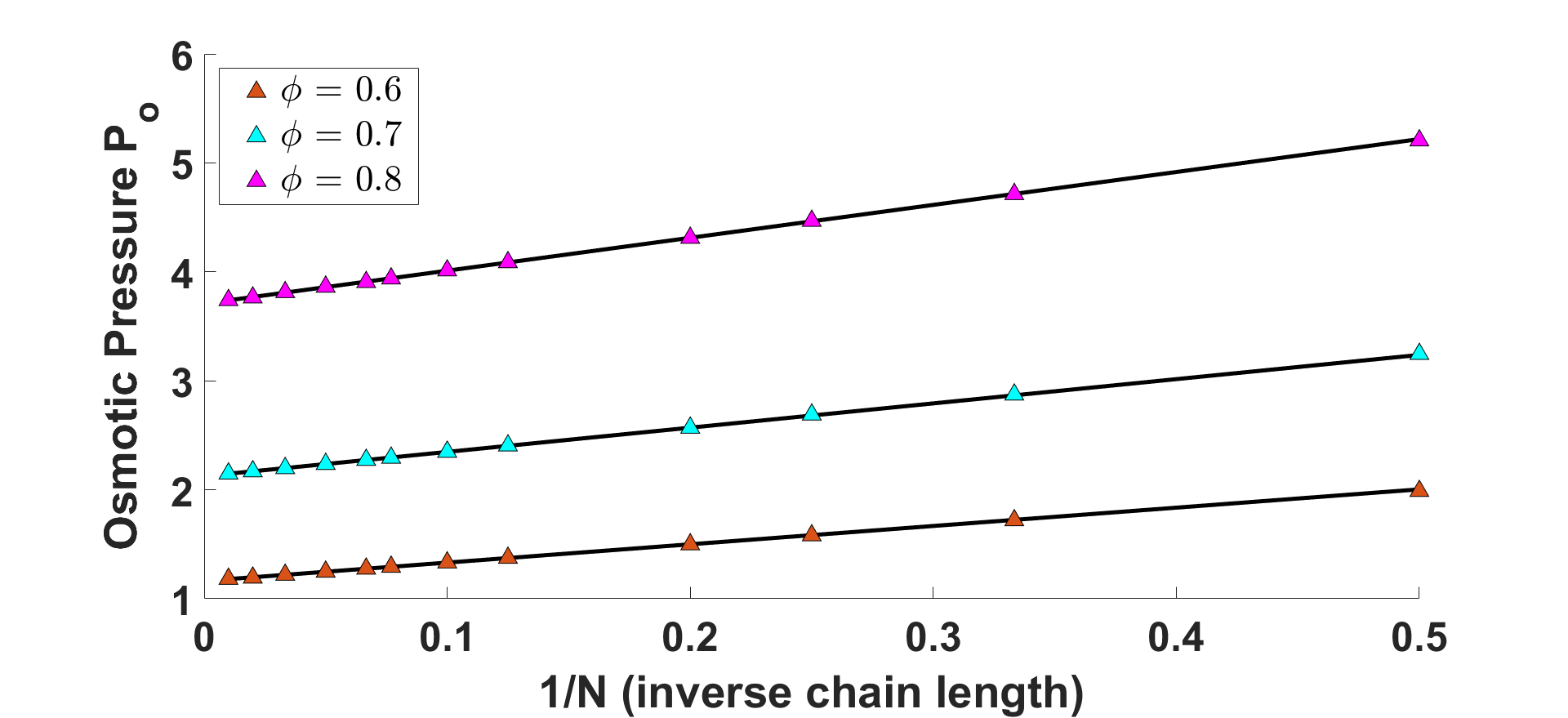}
	\caption{Osmotic pressure vs inverse chain length for densities 0.6 (orange), 0.7 (cyan), and 0.8 (pink).}
	\label{fig:Fig2}
\end{figure}

\begin{figure}[H]
	\centering
	\includegraphics[width=1\linewidth]{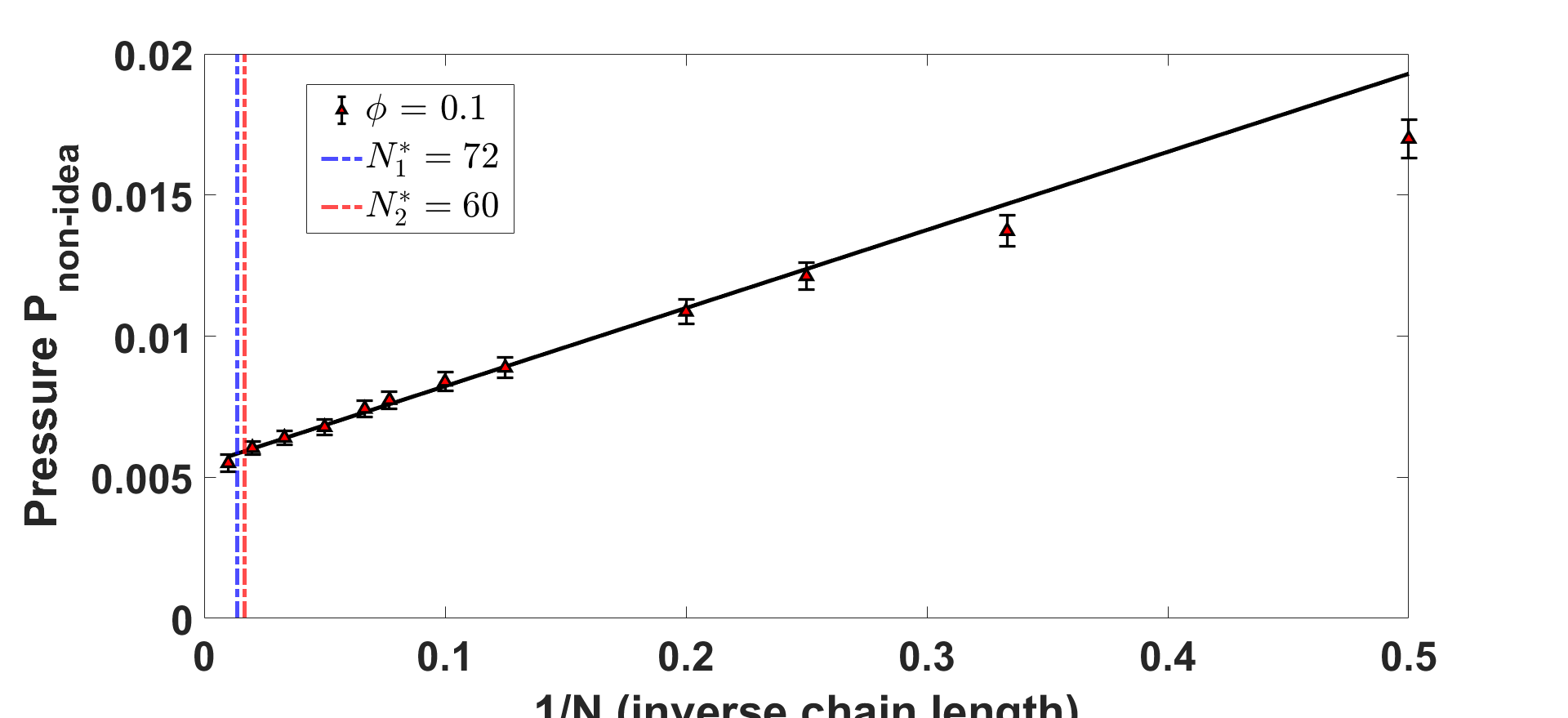}
	\caption{Pressure vs inverse chain length without the ideal gas contribution. A linear dependence on $1/N$ is still observed for relatively large chains. Each point was evaluated 6 times to obtain an average and a standard deviation. The error bars represent the range $[-2\sigma,2\sigma]$ and the points are the averages.}
	\label{fig:Fig3}
\end{figure}

\begin{figure}[H]
	\centering
	\includegraphics[width=1\linewidth]{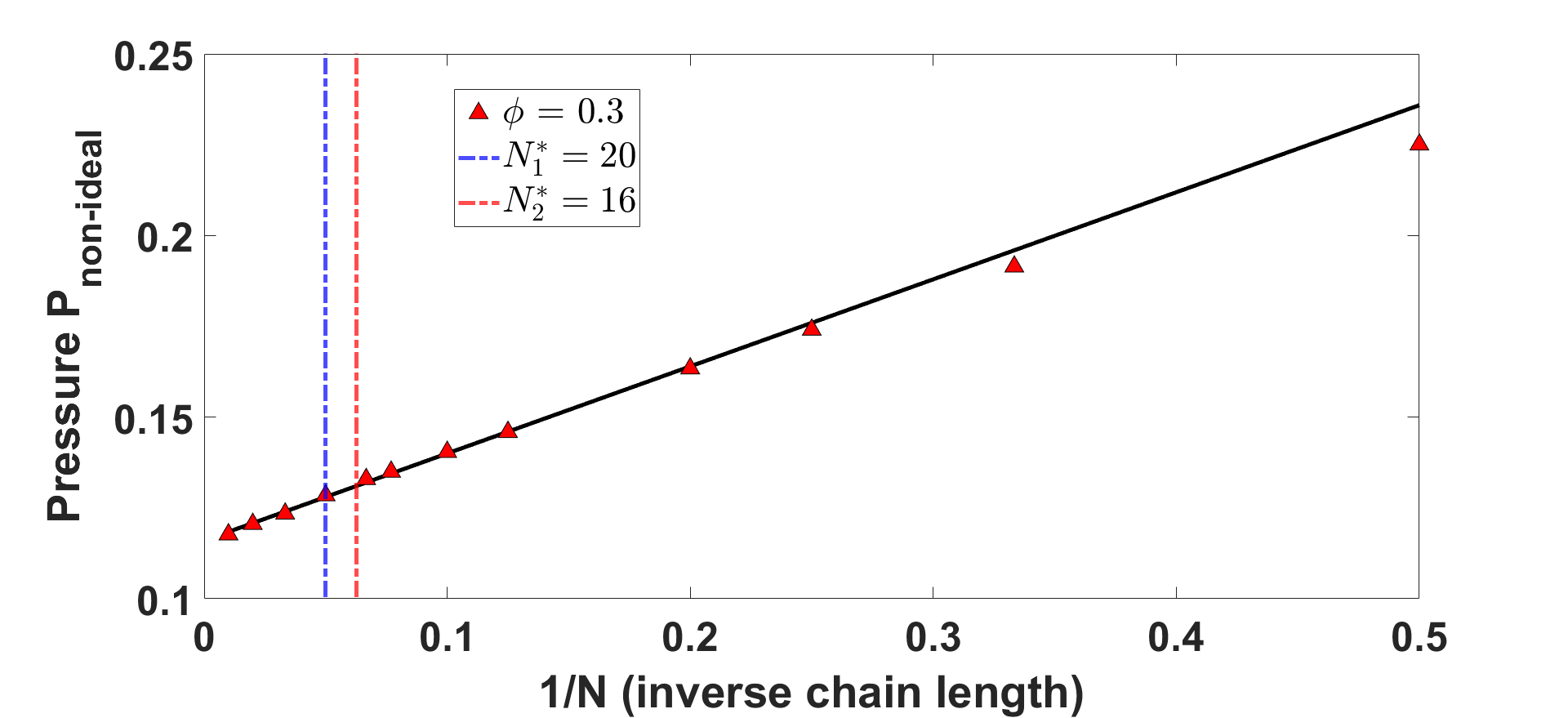}
	\caption{Pressure vs inverse chain length without the ideal gas contribution. A linear dependence on $1/N$ is still observed for relatively large chains.}
	\label{fig:Fig4}
\end{figure}

\begin{figure}[H]
	\centering
	\includegraphics[width=1\linewidth]{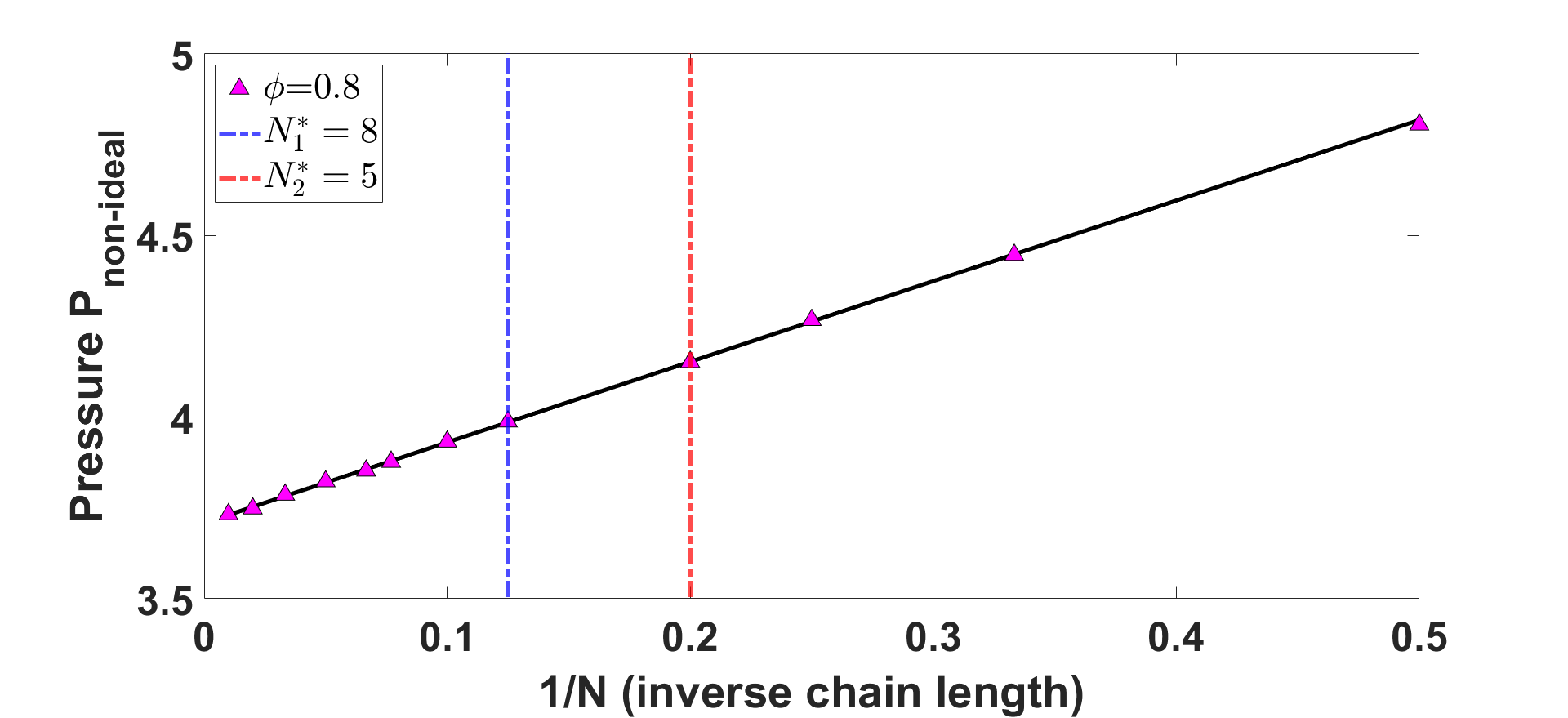}
	\caption{Pressure vs inverse chain length without the ideal gas contribution. A linear dependence on $1/N$ is still observed for relatively large chains.}
	\label{fig:Fig5}
\end{figure}

\begin{figure}[H]
	\centering
	\includegraphics[width=1\linewidth]{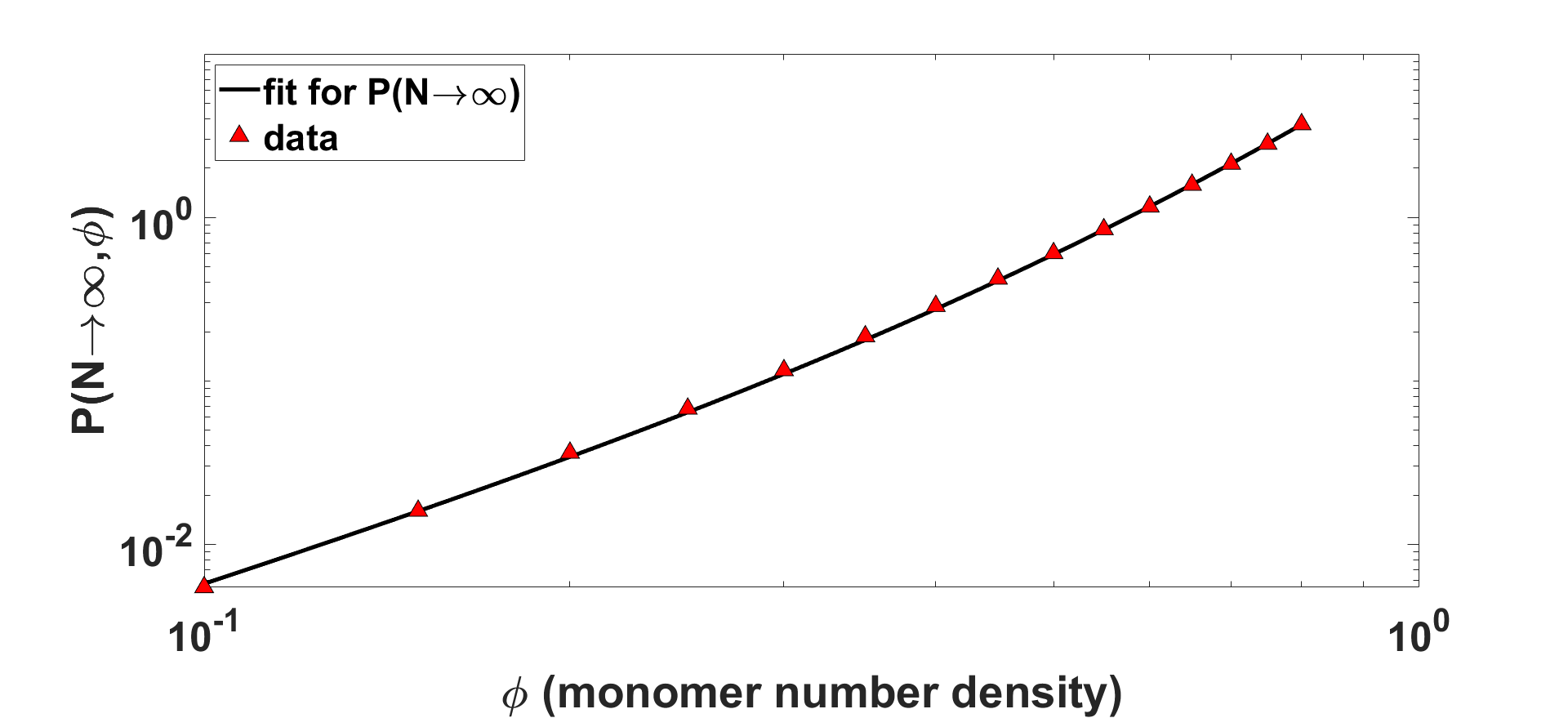}
	\caption{$P(N\rightarrow\infty,\phi)$ vs $\phi$.}
	\label{fig:Fig6}
\end{figure}

\begin{figure}[H]
	\centering
	\includegraphics[width=1\linewidth]{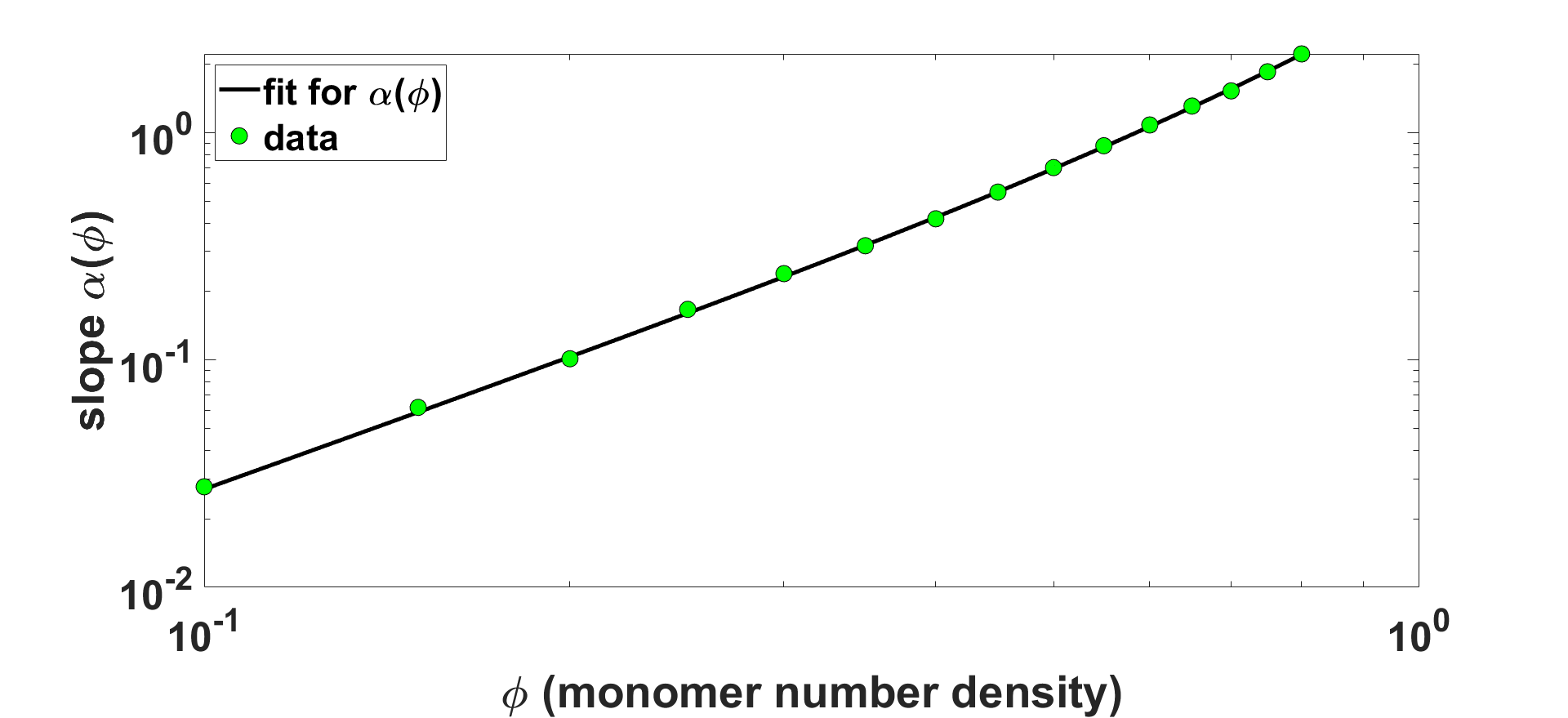}
	\caption{$\alpha(\phi)$ vs $\phi$.}
	\label{fig:Fig7}
\end{figure}

Generally, it is expected that systems in different regimes behave differently, in particular in terms of pressure. To check this, we move to studying the chain length dependence of the osmotic pressure of a monodisperse polymer solution at a fixed monomer number density. Different densities were obtained be keeping the box size and changing the number of chains. The are shown in figures \ref{fig:Fig1} \& \ref{fig:Fig2}. There is a clear linear dependence on $1/N$. This is to be expected when the osmotic pressure is separated into two parts:
\begin{equation}
P(N,\phi)=\frac{\phi}{N}+P_{non-ideal}
\end{equation}
The first part, being the ideal gas contribution to the pressure, indeed does depend linearly on $1/N$, because the density in that part is that of chains, not monomers ($\phi_{chains}=\phi/N$). However, when this contribution is removed, a linear dependence is still remarkably observed, especially for relatively large chains. This is shown in \ref{fig:Fig3}, \ref{fig:Fig4}, \& \ref{fig:Fig5}.
Having this in mind, it is natural to separate $P_{non-ideal}$ into 2 contributions, one being the pressure at infinite $N$ and the other a finite size correction
\begin{equation}
P_{non-ideal}=P(N\rightarrow\infty,\phi)+\alpha(\phi)\frac{1}{N}
\end{equation}
we found that the slope and the y-intercept can be fit fairly well with the following polynomials and are shown in figures \ref{fig:Fig6} and \ref{fig:Fig7}.
\begin{equation}
P(N\rightarrow\infty,\phi)=1.041\phi^{2.3}+5.228\phi^{4}+3.566\phi^{6}
\end{equation}
\begin{equation}
\alpha(\phi)=2.13\phi^{1.9}+1.928\phi^{4}
\end{equation}

\subsection{Binary Mixtures}
\begin{figure}[H]
	\centering
	\includegraphics[width=1\linewidth]{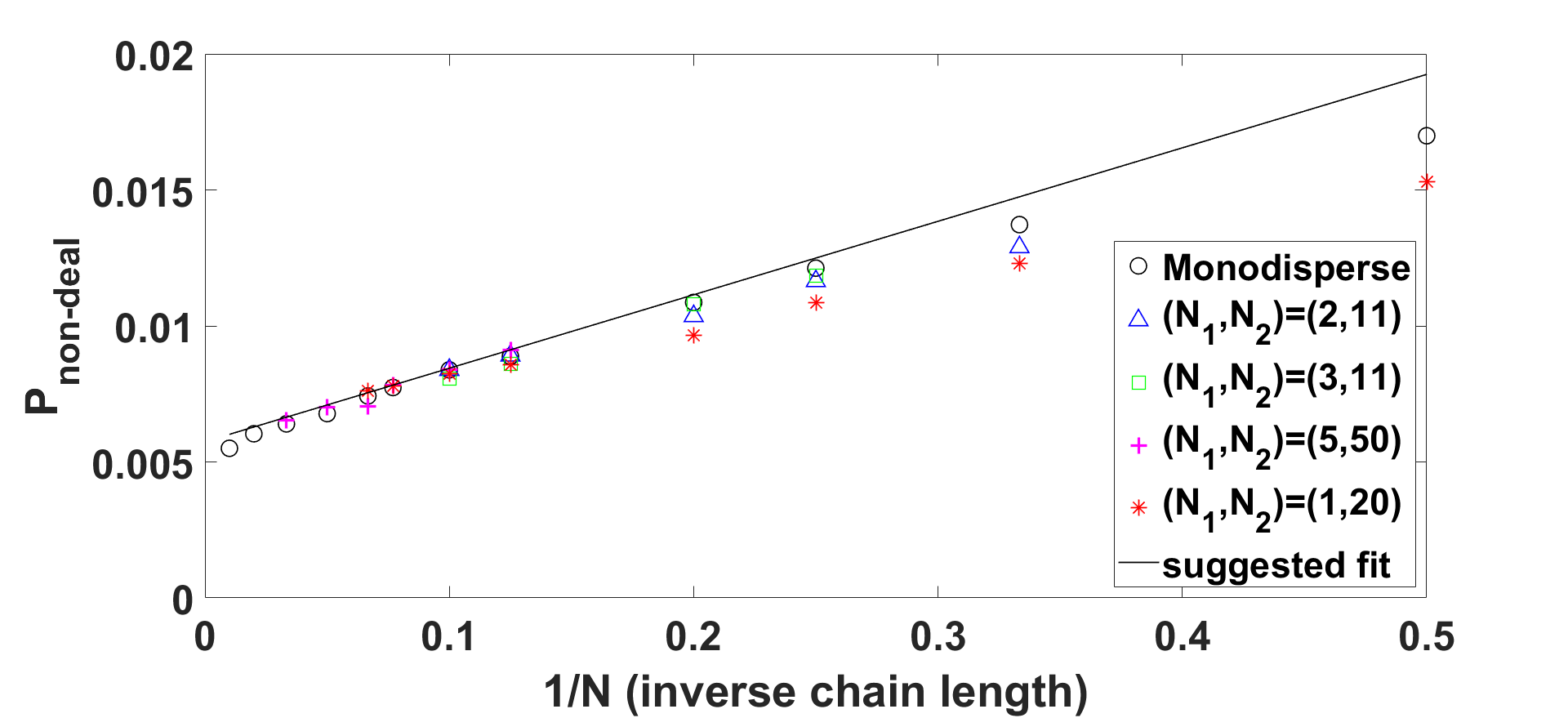}
	\caption{Non-ideal contribution to the osmotic pressure for a monodisperse solution and binary mixtures of lengths with $(N_1,N_2)=[(2,11),(3,11),(5,50),(1,20)]$ for monomer number density 0.1, with the straight line suggested by the phenomenological equation of state.}
	\label{fig:Fig12}
\end{figure}

\begin{figure}[H]
	\centering
	\includegraphics[width=1\linewidth]{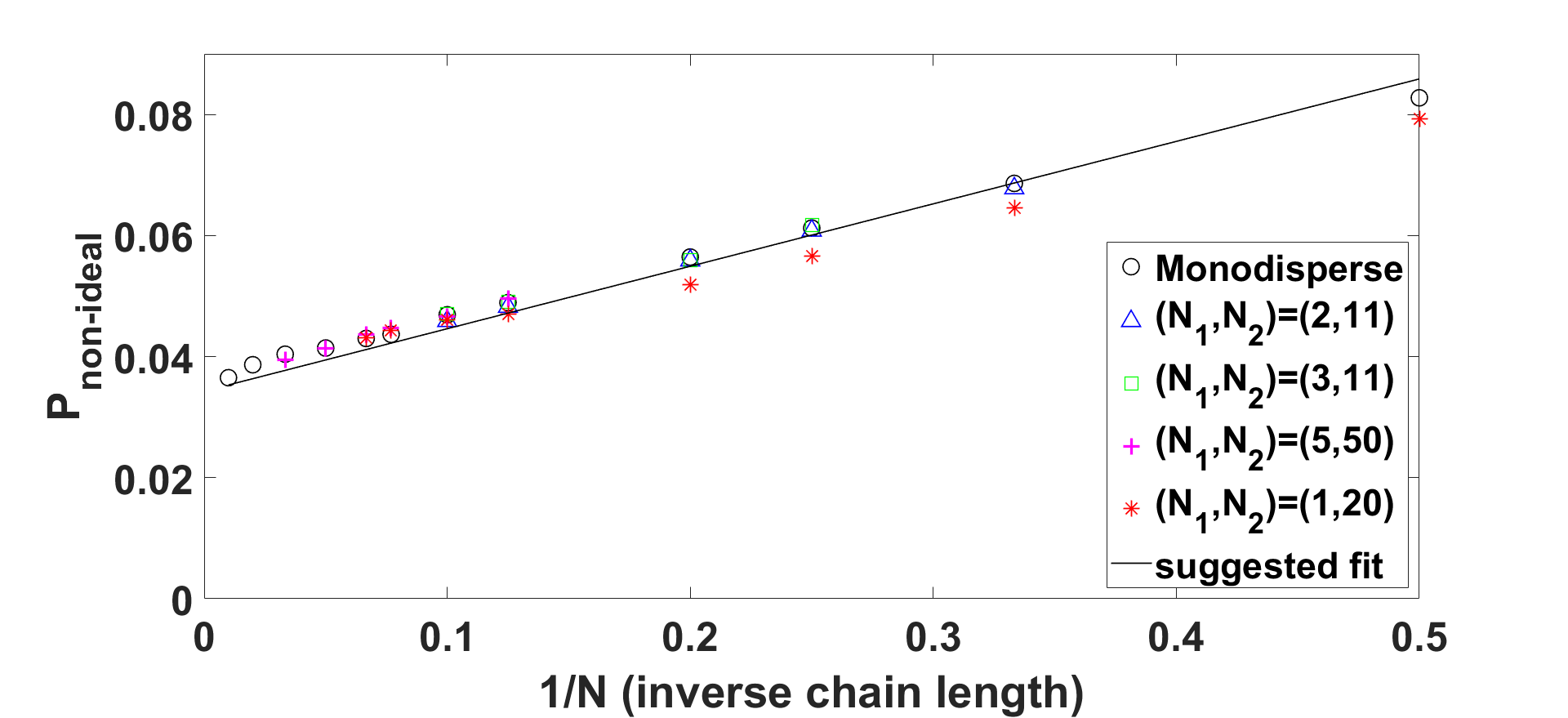}
	\caption{Non-ideal contribution to the osmotic pressure for a monodisperse solution and binary mixtures of lengths with $(N_1,N_2)=[(2,11),(3,11),(5,50),(1,20)]$ for monomer number density 0.2, with the straight line suggested by the phenomenological equation of state.}
	\label{fig:Fig13}
\end{figure}

\begin{figure}[H]
	\centering
	\includegraphics[width=1\linewidth]{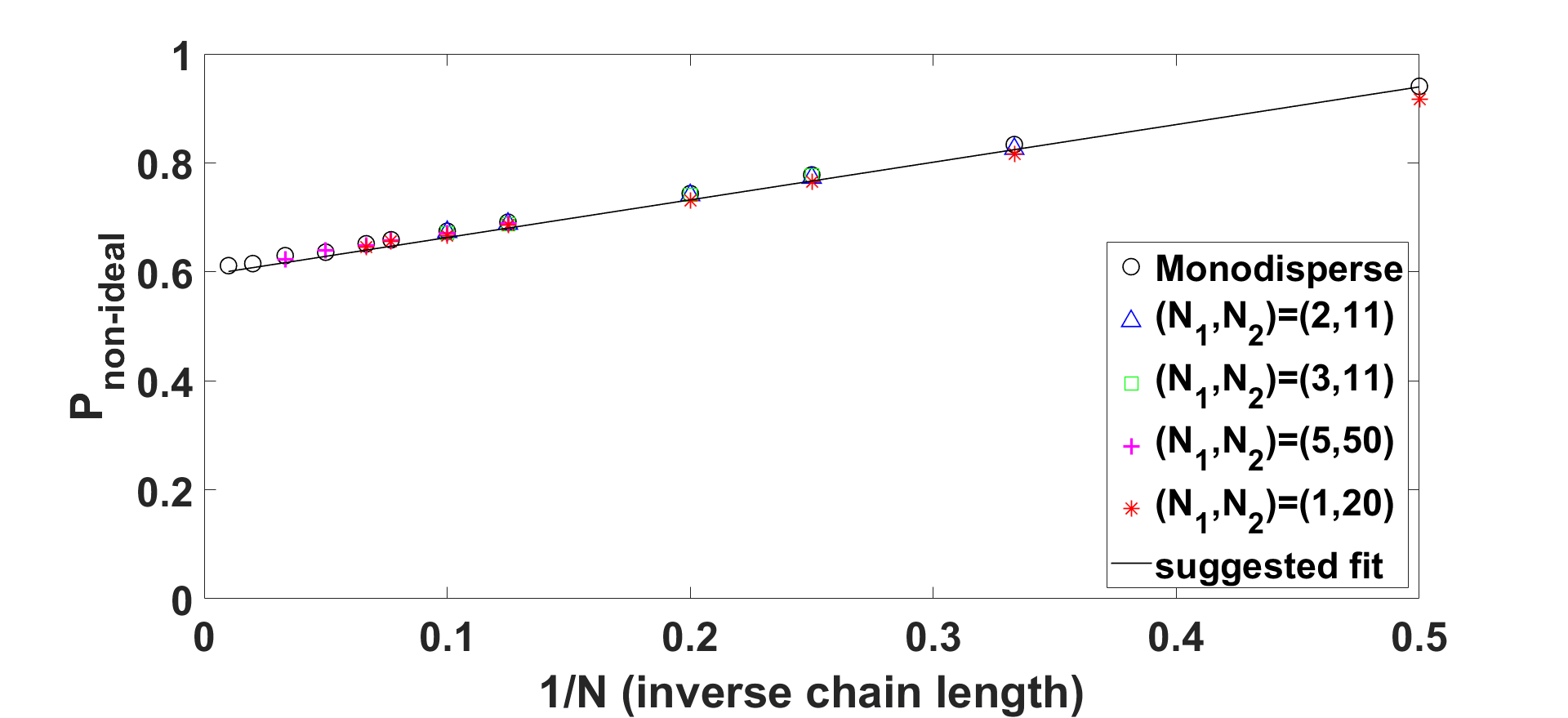}
	\caption{Non-ideal contribution to the osmotic pressure for a monodisperse solution and binary mixtures of lengths with $(N_1,N_2)=[(2,11),(3,11),(5,50),(1,20)]$ for monomer number density 0.5, with the straight line suggested by the phenomenological equation of state.}
	\label{fig:Fig14}
\end{figure}

\begin{figure}[H]
	\centering
	\includegraphics[width=1\linewidth]{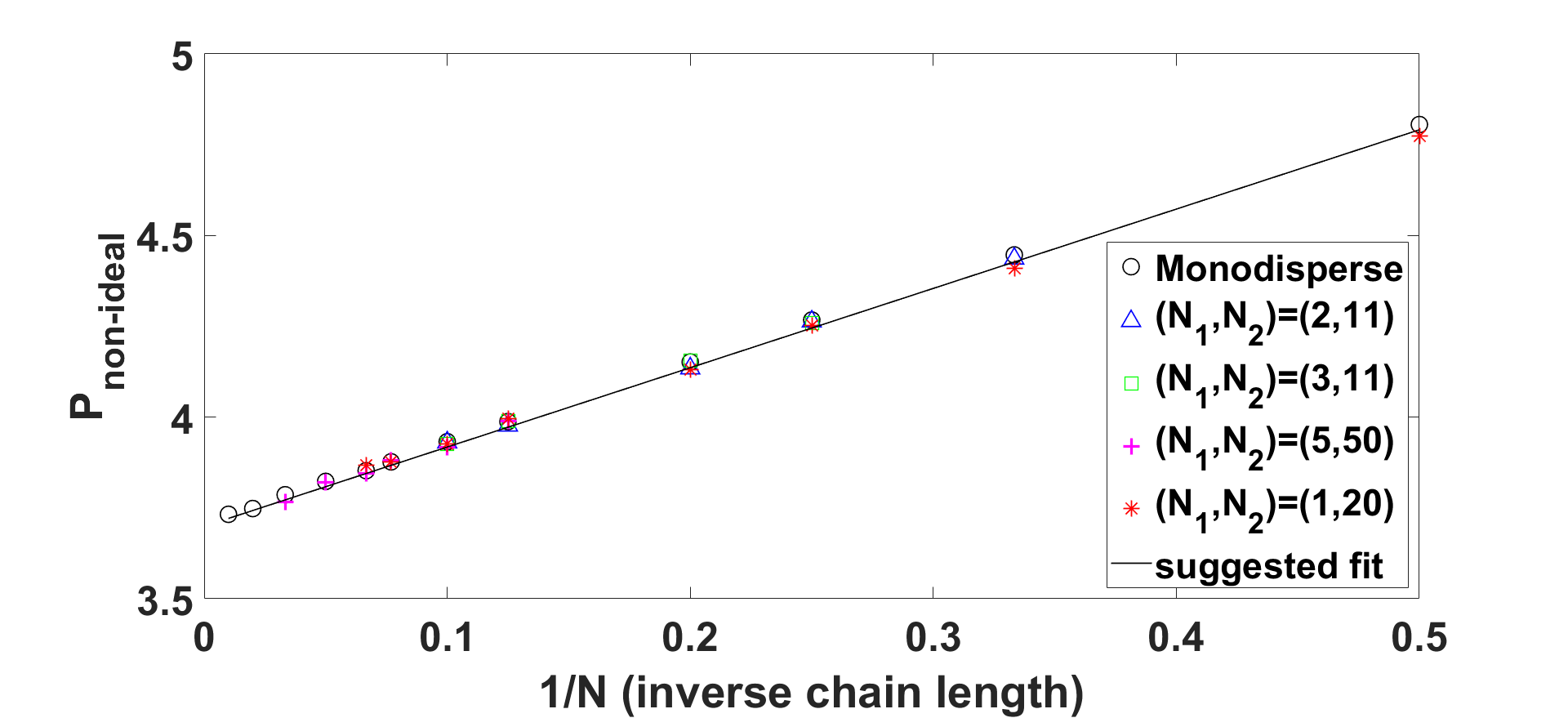}
	\caption{Non-ideal contribution to the osmotic pressure for a monodisperse solution and binary mixtures of lengths with $(N_1,N_2)=[(2,11),(3,11),(5,50),(1,20)]$ for monomer number density 0.8, with the straight line suggested by the phenomenological equation of state.}
	\label{fig:Fig15}
\end{figure}

To see if the above suggested equation of state works for solutions other than monodisperse ones, we check it for binary mixtures, i.e. mixtures of chains of 2 lengths. To do so, we define the inverse effective chain length $1/N_{eff}$ as the weighted sum of the 2 inverse lengths $1/N_1$ and $1/N_2$ with respective monomer number densities $\phi_1$ and $\phi_2$
\begin{equation}
\frac{1}{N_{eff}}=\frac{\phi_1}{\phi}\frac{1}{N_1}+\frac{\phi_2}{\phi}\frac{1}{N_2}
\end{equation}
where $\phi=\phi_1+\phi_2$
The results for different couples $(N_1,N_2)$ and densities are shown in figures \ref{fig:Fig12}, \ref{fig:Fig13}, \ref{fig:Fig14}, and \ref{fig:Fig15}.
\newpage
\bibliographystyle{ieeetr}
\bibliography{references}
\end{document}